\documentclass{ws-p8-50x6-00}
\usepackage{epsfig}

\def\Journal#1#2#3#4{{#1} {\bf #2}, #3 (#4)}

\def\NPA{{\em Nucl. Phys.} A}

\def\PLB{{\em Phys. Lett.}  B}
\def\PRL{\em Phys. Rev. Lett.}

\def\PRC{{\em Phys. Rev.} C}

\def\PPNP{\em Prog. Part. Nucl. Phys.}
\def\EPJA{{\em Eur. Phys. J.} A}

\def\be{\begin{equation}}
\def\ee{\end{equation}}
\def\bea{\begin{eqnarray}}
\def\eea{\end{eqnarray}}

\begin{document}

\title{THERMODYNAMICAL DESCRIPTION OF HEAVY ION COLLISIONS}

\author{T. GAITANOS, H. H. WOLTER}

\address{Sektion Physik, Universit\"at M\"unchen, Am Coulombwall 1\\
D-85748 Garching, Germany\\E-mail: Theo.Gaitanos@Physik.uni-muenchen.de}

\author{C. Fuchs}

\address{Institut f\"ur Theoretische Physik, Universit\"at T\"ubingen, 
Auf der Morgenstelle 14\\D-72076 T\"ubingen, Germany}


\maketitle

\abstracts{We analyze the thermodynamical state of nuclear matter 
in transport descriptions of heavy ion reactions. We determine thermodynamical 
variables from an analysis of local momentum space distributions and compare 
to blast model parameters from an analysis of fragment energy spectra. These 
descriptions are applied to spectator and fireball matter in semi-central and 
central Au+Au collisions at SIS-energies, respectively.}

A topic of great interest in the study of intermediate energy heavy-ion collisions 
is the investigation of the nuclear equation of state away from saturation at 
lower densities and at finite temperature in connection to possible signals of 
phase transitions \cite{rw00,po97,ala00,hir99}. Theoretical multifragmentation 
models \cite{stat} have addressed this question , but their application in the dynamical 
situation of heavy-ion collisions is difficult due to non-equilibrium \cite{gait99} 
and finite size effects \cite{statfinite}. The question is whether the nuclear 
matter in heavy-ion collisions is both in thermodynamical equilibrium and instable, 
i.e. whether the fragment emitting source can be characterized by 
a thermodynamically well defined freeze-out configuration. 
In this work we investigate this question by studying the results of transport 
calculations described in detail in Refs. \cite{temp,gait00}. 
We have applied thermodynamical analyses to spectator and participant matter 
using either the information of the local 
momentum distribution to define a local thermodynamical temperature ($T_{loc}$) or 
from a blast model analysis \cite{reisdorf2} of fragment energy spectra to obtain a 
slope parameter ($T_{slope}$) of the fragment spectra. We compare these 
``theoretical data'' with experiments.

\begin{figure}[t]
\begin{center}
\unitlength1cm
\begin{picture}(12,3.2)
\put(0,-0.5){\makebox{\psfig{file=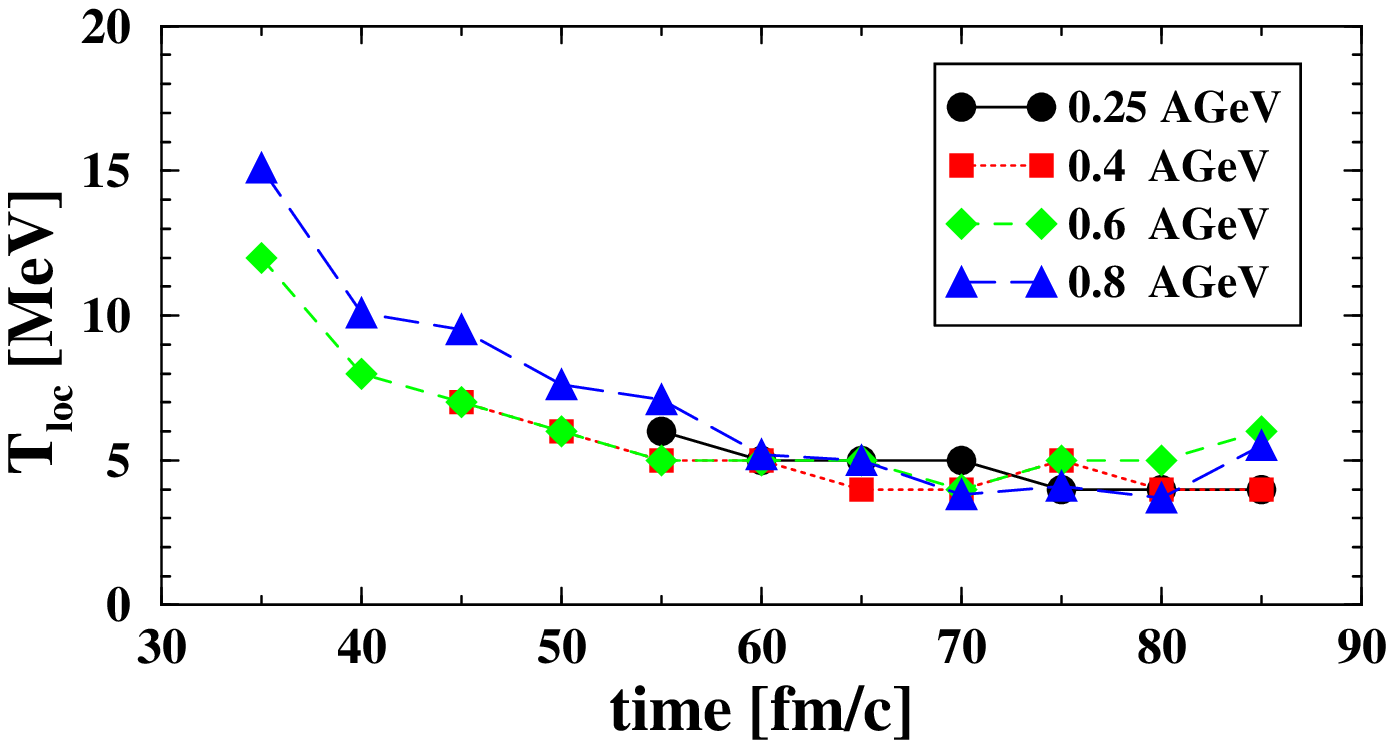,width=6.0cm}}}
\put(6,-0.5){\makebox{\psfig{file=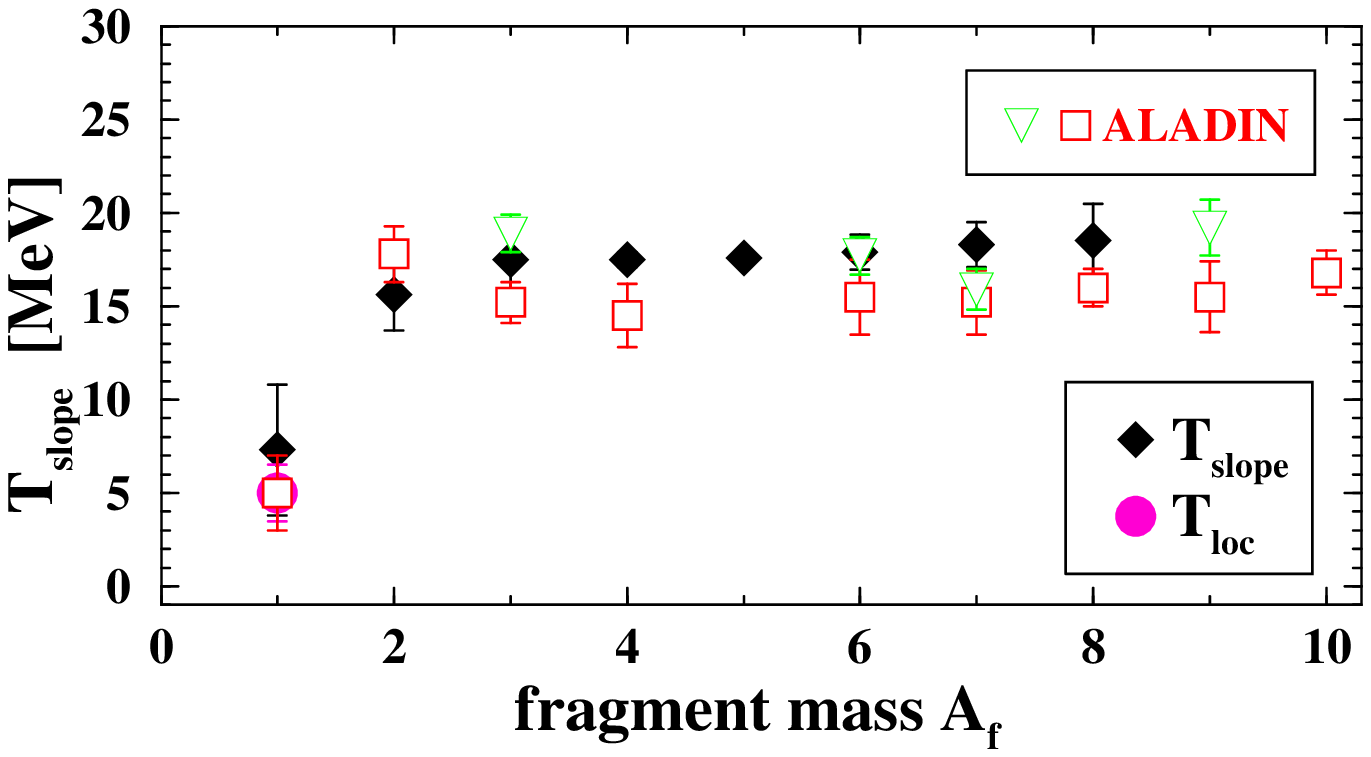,width=6.0cm}}}
\end{picture}
\caption{\label{fig1}Left: local temperatures $T_{loc}$ as function of time 
in spectators for semi-central Au+Au reactions. 
Right: slope temperatures $T_{slope}$ for spectator fragments as function of $A_{f}$ with 
comparison to ALADIN data \protect\cite{ala00}. 
The local temperature of nucleons $T_{loc}$ is also shown.}
\end{center}
\end{figure}
Fig.~\ref{fig1} on the left shows the temporal evolution of the local temperature 
$T_{loc}$ in spectator matter for semicentral ($b=4.5$ fm) Au+Au reactions at different 
energies, as obtained from fits to 
the local momentum space distribution of transport calculations. After 
the spectators are well defined $T_{loc}$ approaches rather constant values 
and remains fairly 
stable for several fm/c. The local temperature is independent on the incident energy 
indicating the existence of an intermediate equilibrium state. By considering 
an effective compressibility $K \sim \partial P / \partial \rho$ \cite{temp} and 
instability by $K<0$ we find that after 
$t \stackrel{>}{\sim} 40-45$ fm/c the spectator enterns into an instability region 
\cite{temp} and therefore may break up into fragments. The breakup conditions of 
$T_{loc} \sim 5-6$ MeV and $\rho \sim (\frac{1}{3}-\frac{1}{2}) \rho_{sat}$ are 
consistent with experimental isotope temperatures and densities extracted from 
$2$-particle correlations \cite{ala00,schwarz}. The generation of fragments by 
a phase space coalescence 
in the spectator matter and an analysis of the fragment energy 
spectra in the blast model \cite{reisdorf2} 
leads to the results of Fig.~\ref{fig1} on the right. The independence of 
$T_{slope}$ on $A_{f}$ indicates that the fragments are emitted from an equilibrated 
source. However, the slope temperature of the fragments is higher compared to $T_{slope}$ 
of the nucleons, which in turn is equal to $T_{loc}$. 
This is understood by the Fermi-motion of the nucleons in the fragmenting source in the 
spirit of the Goldhaber model, 
as discussed in Refs. \cite{ala00,gait00,gold,bauer}. The slope parameter of the fragments is 
thus consistent with the nucleon $T_{slope}$ and also with the local temperature 
$T_{loc}$. We thus conclude the existence of a common freeze-out configuration 
for nucleons and fragments in spectator decay. This interpretation is also supported 
by the experiments \cite{ala00} shown in the Fig.~\ref{fig1}.

\begin{figure}[t]
\begin{center}
\unitlength1cm
\begin{picture}(12,5.5)
\put(2.,-0.5){\makebox{\psfig{file=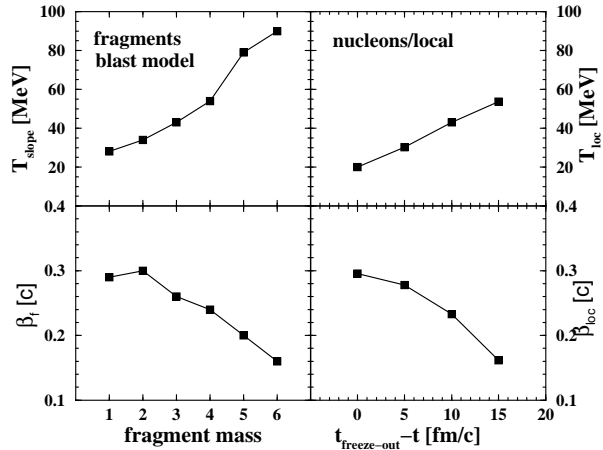,width=8.0cm}}}
\end{picture}
\caption{\label{fig2}(left) Slope temperatures and radial flow as function of fragment 
mass. (right) Local values from the momentum distributions at times before the freeze-out 
for a central Au+Au reaction at $0.6$ AGeV.}
\end{center}
\end{figure}
The results of the same analysis for participant matter in central heavy-ion reactions are 
summarized in Fig.~\ref{fig2} where a radial flow velocity $\beta_{f}$ appears as an 
additional parameter in the blast model fits due to the radial expansion of the fireball. 
We observe that slope temperatures rise and flow velocities fall with increasing fragment 
mass in contrast to the behavior for spectator fragments in Fig.~\ref{fig1}. A 
similar behavior has been seen experimentally at $1$ AGeV in \cite{eos} and 
theoretically in \cite{daffin,hombach}. This behavior cannot be interpreted as 
fragments originating from a common freeze-out state, i.e. from a unique fragmenting 
source. To arrive at an interpretation 
we have compared these results with the local temperatures and flow velocities for 
different times before the nucleon freeze-out, i.e. for $t^{\prime}=t_{freeze-out}-t$ 
with $t_{freeze-out} \sim 35$ fm/c. It is seen that for $A_f = 1$ the values at 
freeze-out are close to the blast model ones, as required. However, for fragment 
masses $A_f > 1$ the slope temperatures and velocities behave qualitatively very 
similar to the local temperatures and flow velocities at earlier times. This would 
suggest to interpret the fragment temperatures and velocities as signifying that heavier 
fragments originate at times earlier than the nucleon freeze-out. This may not be 
unreasonable since in order to make a heavier fragment one needs higher densities 
which occur at earlier times and hence higher temperatures. However, this does not 
neccessarily imply that the fragments are really 
formed at this time, since fragments could hardly survive such high temperatures, 
as also discussed in refs. \cite{reisdorf2,botv}. But it could mean that these 
fragments carry information about this stage of the collision. In any 
case it means that in the participant region fragments are {\it not} 
formed in a common equlibrated freeze-out configuration, and that in 
such a situation slope temperatures have to be interpreted with great caution.

In summary fragmentation phenomena in heavy ion collisions are studied as a 
means to explore 
the phase diagram of hadronic matter. For this it is neccessary to determine the 
thermodynamical properties of the fragmenting source. One way to do this 
experimentally is to investigate fragment kinetic energy spectra. In theoretical 
simulations the thermodynamical state can be obtained locally in space and time 
from the phase space distribution. In this work we have compared this with the 
information obtained from the generated fragment spectra. We apply this method 
to the spectator and participant regions of relativistic Au+Au-collisions. 
We find that the spectator represents a well developed, equilibrated  and 
instable fragmenting source. 
In the participant region the local temperature at the 
nucleon freeze-out and the slope temperatures from fragment spectra behave 
differently from those of the spectator. 
The slope temperatures rise with fragment mass which might indicate 
that the fragments are not formed in a common, equilibrated source. 



\begin{thebibliography}{99}

\bibitem{rw00}
J. Richert, P. Wagner, nucl-th/0009023 (to appear in Phys. Rep.).

\bibitem{po97}
J. Pochodzalla, \Journal{\PPNP}{39}{443}{1997}.

\bibitem{ala00}
T. Odeh et al., \Journal{\PRL}{84}{4557}{2000}.

\bibitem{hir99}
Proc. of the International 
Workshop XXVII on Gross Properties of Nuclei and Nuclear 
Excitations, Hirschegg, Austria, 1999.

\bibitem{stat}
J.P. Bondorf et al., \Journal{\NPA}{443}{321}{1985};\\
A. S. Botvina, D.H.E. Gross, \Journal{\NPA}{592}{257}{1995}  and contribution to this 
conference.


\bibitem{gait99}
H.H. Wolter, C. Fuchs, T. Gaitanos, \Journal{\PPNP}{42}{137}{1999};\\
T. Gaitanos, C. Fuchs, H.H. Wolter, \Journal{\NPA}{650}{97}{1999}.

\bibitem{statfinite}
PH. Chomaz, F. Gulminelli, \Journal{\NPA}{647}{153}{1999}; \\
M. D'Agostino et al., \Journal{\PLB}{473}{219}{2000} and contribution to this 
conference.

\bibitem{temp}
C. Fuchs, P. Essler, T. Gaitanos, H.H. Wolter, \Journal{\NPA}{626}{987}{1997}.

\bibitem{gait00}
T. Gaitanos, H.H. Wolter, C. Fuchs \Journal{\PLB}{478}{79}{2000}.

\bibitem{reisdorf2}
P.J. Siemens, J.O. Rasmussen, \Journal{\PRL}{42}{880}{1979};\\
W. Reisdorf et al., \Journal{\NPA}{612}{493}{1997}.

\bibitem{schwarz}
S. Fritz et al., \Journal{\PLB}{461}{315}{1999}.

\bibitem{gold}
A. S. Goldhaber, \Journal{\PLB}{53}{306}{1974}; 
\Journal{\PRC}{17}{2243}{1978}.

\bibitem{bauer}
W. Bauer, \Journal{\PRC}{51}{803}{1995}.

\bibitem{eos}
M. Lisa et al., \Journal{\PRL}{75}{2662}{1995}.

\bibitem{daffin}
F. Daffin, K. Haglin,  W. Bauer, \Journal{\PRC}{54}{1375}{1996}.

\bibitem{hombach}
A. Hombach et al., \Journal{\EPJA}{5}{157}{1999}.

\bibitem{botv}
W. Neubert, A.S. Botvina, \Journal{\EPJA}{7}{101}{2000}.

\end{thebibliography}
\end{document}